\documentclass[fleqn,usenatbib]{mnras}

\usepackage[T1]{fontenc}

\DeclareRobustCommand{\VAN}[3]{#2}
\let\VANthebibliography\thebibliography
\def\thebibliography{\DeclareRobustCommand{\VAN}[3]{##3}\VANthebibliography}

\usepackage{graphicx}	
\usepackage{amsmath,stmaryrd}	
\usepackage{tikz}
\usepackage{CJK}
\usepackage{multirow}
\usepackage{algorithmicx}
\usepackage{algpseudocode}
\usepackage[caption=false]{subfig}
\usepackage{amssymb,amstext}
\usepackage{enumitem}

\usepackage{newtxtext,newtxmath}

\newcommand{\half}{\tfrac{1}{2}}
\newcommand{\p}{\partial}

\newcommand\llangle{\langle\!\langle}
\newcommand\rrangle{\rangle\!\rangle}

\def\bj{{\bf j}}
\def\be{{\bf e}}
\def\br{{\bf r}}
\def\bR{{\bf R}}

\def\br{{\bf r}}

\def\bR{{\bf R}}

\def\bq{{\bf q}}
\def\bp{{\bf p}}

\def\df{{\caps df}}

\def\half{\tfrac{1}{2}}

\def\br{{\bf r}}
\def\bx{{\bf x}}

\def\br{{\bf r}}

\newcommand{\jhat}{\boldsymbol{\hat{\textbf{\j}}}}

\newfont{\caps}{cmcsc10}

\title[Hamiltonian for ZLK oscillations]{The Hamiltonian for von Zeipel--Lidov--Kozai oscillations}

\author[Tremaine]{
Scott Tremaine$^{1,2}$\thanks{E-mail: tremaine@ias.edu}
\\
$^{1}$Canadian Institute for Theoretical Astrophysics, University of Toronto, 60 St. George Street, Toronto, ON M5S 3H8, Canada\\
$^{2}$School of Natural Sciences, Institute for Advanced Study, Princeton, NJ 08540, USA
}

\date{Accepted XXX. Received YYY; in original form ZZZ}

\pubyear{2023}

\begin{document}
\label{firstpage}
\pagerange{\pageref{firstpage}--\pageref{lastpage}}
\maketitle

\begin{abstract}

The Hamiltonian used in classical analyses of von Zeipel--Lidov--Kozai or ZLK oscillations in hierarchical triple systems is based on the quadrupole potential from a distant body on a fixed orbit, averaged over the orbits of both the inner and the outer bodies (``double-averaging''). This approximation can be misleading, because the corresponding Hamiltonian conserves the component of angular momentum of the inner binary normal to the orbit of the outer binary, thereby restricting the volume of phase space that the system can access. This defect is usually remedied by including the effects of the octopole potential, or by allowing the outer orbit to respond to variations in the inner orbit. However, in a wide variety of astrophysical systems nonlinear perturbations are comparable to or greater than these effects. The long-term effects of nonlinear perturbations are described by an additional Hamiltonian, which we call Brown's Hamiltonian. At least three different forms of Brown's Hamiltonian are found in the literature; we show that all three are related by a gauge freedom, although one is much simpler than the others. We argue that investigations of ZLK oscillations in triple systems should include Brown's Hamiltonian. 

\end{abstract}


\section{Introduction}

\label{sec:intro}

von Zeipel--Lidov--Kozai or ZLK oscillations, sometimes called Lidov--Kozai or Kozai oscillations, occur in hierarchical triple systems such as a star that is orbited by a planet and a distant companion star, a planet and a satellite orbiting their host star, a triple-star system, or a binary star orbiting a supermassive black hole. In these systems the distant body can excite large oscillations in the eccentricity and inclination of the inner binary. A remarkable feature of ZLK oscillations is that large oscillations can be excited by a third body of arbitrarily small mass or at arbitrarily large distance: as its mass shrinks or the distance grows, the period of the oscillations increases but their amplitude remains approximately the same. ZLK oscillations were originally discovered by \nocite{zeipel10} von Zeipel (1910; see also \citealt{ito19}), re-discovered by \cite{lidov62}, and popularized in the West by \cite{kozai62}. For reviews, see \cite{naoz16} and \cite{shev17}.

In the simplest treatment of ZLK oscillations, the gravitational potential from the distant body is approximated by (i) keeping only the quadrupole terms in the potential, (ii) ignoring any variations in the outer orbit, which is usually legitimate since the outer orbit has much more angular momentum than the inner one; (iii) averaging the Hamiltonian based on the quadrupole potential over the orbits of both the inner and the outer binary, an approximation sometimes called ``double-averaging''. We shall call a theory that uses these approximations ``classical'' ZLK theory. The perturbing Hamiltonian in classical ZLK theory is given below by equation (\ref{eq:quadd}).

Let $a_1$ and $a_2$ be the semimajor axes of the inner and outer binary, $m_0$ and $m_1$ the masses of the bodies in the inner binary, and $m_2$ the mass of the distant body. Then the typical gravitational potential in the inner binary is $\Phi_\mathrm{K}\sim G(m_0+m_1)/a_1$ (``K'' for ``Kepler'') and the quadrupole potential of the distant body is $\Phi_\mathrm{quad}\sim Gm_2a_1^2/a_2^3$. Thus the fractional strength of the perturbation in classical ZLK theory is
\begin{equation}
  \epsilon^2 \sim \frac{\Phi_\mathrm{quad}}{\Phi_\mathrm{K}}\sim
  \frac{m_2}{m_0+m_1}\left(\frac{a_1}{a_2}\right)^3. 
  \label{eq:epsdef1}
\end{equation}
When $m_2\gtrsim m_0+m_1$ this result can be expressed simply in terms of the mean motions $n_1$ and $n_2$ of the inner and outer binaries:
$n_1^2=G(m_0+m_1)/a_1^3$, $n_2^2=G(m_0+m_1+m_2)/a_2^3 \sim
Gm_2/a_2^3$, so
\begin{equation}
  \epsilon \sim \frac{n_2}{n_1}.
  \label{eq:epsdef2}
\end{equation}

An accidental property of classical ZLK theory is that the perturbing Hamiltonian (\ref{eq:quadd}) is independent of the longitude of the node of the inner binary on the outer orbit, $\Omega_1$, even when the outer orbit is eccentric. Since the longitude of node is the coordinate conjugate to the $z$-component of the angular momentum of the inner binary, $L_{z1}$ (eq.\ \ref{eq:delaunay}), we conclude that $L_{z1}$ is conserved. As a result, the Hamiltonian for classical ZLK theory is autonomous with only one degree of freedom: a single angle $\omega_1$, the argument of periapsis, and the conjugate action $L_1$, the angular momentum of the inner binary (eq.\ \ref{eq:delaunay}). Therefore the trajectories in classical ZLK theory are integrable, and in fact can be expressed analytically as Jacobian elliptic functions \citep{kn99}.

The integrability of classical ZLK theory is both a feature and a bug -- a feature because it enables the properties of ZLK oscillations to be derived analytically, but a bug because it restricts the motion to a two-dimensional manifold in phase space (with coordinates $\omega_1$ and $L_{z1}$) and suppresses important dynamical behavior such as chaos that can only appear in systems with more than one degree of freedom. As an example, suppose that the two bodies in the inner binary are initially on a circular orbit, with radius much larger than the physical radius of either body. Then since $L_{z1}$ is conserved the bodies cannot collide unless $L_{z1}$ is nearly zero, which in turn requires that the initial inclination between the inner and outer binaries is near $90^\circ$. In contrast, more realistic ZLK oscillations can lead to collisions from a wide range of initial inclinations.

These considerations motivate more accurate analyses than classical ZLK theory. One approach is to add octopole terms to the gravitational potential from the distant body, while continuing to average over both the inner and outer orbits; this is sometimes called ``eccentric'' ZLK theory since the orbit-averaged octopole potential vanishes unless the eccentricity of the outer orbit is non-zero, as seen in equation (\ref{eq:fordo}). The octopole potential is $\Phi_\mathrm{oct}\sim Gm_2a_1^3/a_2^4$ and the fractional strength of the octopole perturbation is
\begin{equation}
  \frac{\Phi_\mathrm{oct}}{\Phi_\mathrm{K}}\sim \frac{m_2}{m_0+m_1}\left(\frac{a_1}{a_2}\right)^4,
  \label{eq:octoom}
\end{equation}
smaller than $\epsilon^2$ by a factor of $a_1/a_2$. 

A second approach is to allow the outer orbit to respond to variations in the inner orbit. The perturbations in the outer orbit are typically smaller than those in the inner orbit by the ratio of the angular momenta $L_1/L_2$, so from equation (\ref{eq:jrat}) the fractional strength of these effects is
\begin{equation}
    \frac{\Phi_\mathrm{quad}}{\Phi_\mathrm{Kep}}\frac{L_1}{L_2}\sim \frac{m_0m_1(m_0+m_1+m_2)^{1/2}}{(m_0+m_1)^{5/2}}\frac{a_1^{7/2}}{a_2^{7/2}}\frac{(1-e_1^2)^{1/2}}{(1-e_2^2)^{1/2}}.
\end{equation}
Of course these effects are negligible in the common situation when $m_1\ll m_0,m_2$ (e.g., a satellite orbiting a planet).

Another route to higher accuracy is to include the effects of nonlinear quadrupole perturbations to the orbit of the inner binary. The quadrupole Hamiltonian due to the distant body can be written schematically as a sum of terms of the form $A_{k_1k_2}\exp[i(k_1\lambda_1+k_2\lambda_2)]$ where $\lambda_1$ and $\lambda_2$ are the mean longitudes of the inner and outer binaries, $k_1$ and $k_2$ are integers, and $A_{k_1k_2}$ is a function of the actions and other angles that is of order $\epsilon^2$ relative to the Kepler Hamiltonian. The terms with $k_1=0$ correspond to the average of the Hamiltonian over the orbit of the inner binary, sometimes called the ``single-averaged'' Hamiltonian. The terms with $k_1=k_2=0$ correspond to the double-averaged Hamiltonian. In canonical perturbation theory, the periodic terms -- the complement of the terms in the double-averaged Hamiltonian -- are suppressed by a canonical transformation to new actions and angles that differ from the old ones by fractions of order $\epsilon^2n_1/(k_1n_1+k_2n_2)$ where $n_1$ and $n_2$ are the mean motions. This canonical transformation gives rise to a new Hamiltonian that contains terms independent of the mean longitudes that are of order $\epsilon^4n_1/(k_1n_1+k_2n_2)$ relative to the Kepler Hamiltonian. The largest of these have $k_1=0$ since $n_1\gg n_2$, which means that the strongest nonlinear effects can be derived from the single-averaged Hamiltonian. The fractional strength of this perturbation is therefore
\begin{equation}
   \epsilon^4
  \frac{n_1}{n_2}\sim \frac{m_2^2}{(m_0+m_1)^{3/2}(m_0+m_1+m_2)^{1/2}}\left(\frac{a_1}{a_2}\right)^{9/2}.
\end{equation}
When $m_2\gtrsim m_0+m_1$ this expression simplifies to 
\begin{equation}
  \frac{n_2^3}{n_1^3}\sim \epsilon^3.
  \label{eq:brownoom}
\end{equation}
One of the subtleties of hierarchical systems is that although the effects of this order arise from second-order perturbations, their amplitude is $\sim\epsilon^3$ rather than $\epsilon^4$. The Hamiltonian derived in this way is sometimes called the ``second-order'' Hamiltonian, even though it is of order $\epsilon^3$; to avoid confusion we shall call it Brown's Hamiltonian in honor of E. W.\ Brown, who first investigated its properties at arbitrary inclinations and eccentricities \citep{brown1,brown2,brown3}. Brown's Hamiltonian is given in equation (\ref{eq:hbrown}).  In many hierarchical triple systems the effects of Brown's Hamiltonian are comparable to or stronger than the effects of the octopole Hamiltonian or variations in the outer orbit, so it should be included in analyses of ZLK oscillations that aim to improve on the classical one. The main goal of this paper is to derive Brown's Hamiltonian and explore some of its properties.

Historically, these issues were first investigated in the attempt to understand the average rate of precession of the lunar longitudes of periapsis and node, $\dot\varpi_1$ and $\dot\Omega_1$, in the triple system consisting of the Earth (body 0), the Moon (body 1) and the Sun (body 2). If the octopole potential is neglected, and the lunar eccentricity and inclination to the ecliptic are assumed to be small, we have \citep{bc61}
\begin{align}
  \frac{1}{n_1}\frac{d\varpi_1}{d t}&=\tfrac{3}{4}\epsilon^2+\tfrac{225}{32}\epsilon^3+\tfrac{4\,071}{128}\epsilon^4 +\tfrac{265\,493}{2\,048}\epsilon^5+\mbox{O}(\epsilon^6), \nonumber \\
 \frac{1}{n_1}\frac{d\Omega_1}{d t}&=-\tfrac{3}{4}\epsilon^2+\tfrac{9}{32}\epsilon^3+\tfrac{273}{128}\epsilon^4+\tfrac{9\,797}{2\,048}\epsilon^5+\mbox{O}(\epsilon^6).
 \label{eq:bc}
\end{align}
In this equation only, $\epsilon$ is exactly equal to $n_2/n_1$, which is 0.0748 for the Earth--Moon--Sun system. The terms proportional to $\epsilon^2$ correspond to classical ZLK theory and those proportional to $\epsilon^3$ arise from Brown's Hamiltonian\footnote{This is shown explicitly in the Appendix.}. The series for $\dot\varpi$ is given to $\epsilon^{11}$ by \cite{hill94} and the series for $\dot\Omega$ is given to $\epsilon^6$ by \cite{delaunay60,delaunay67}. 

The terms of order $\epsilon^2$ were known to Newton. Unfortunately the series for $\dot\varpi$ is notoriously slow to converge: the value of $\dot\varpi$ obtained from the first term in the series is smaller than the exact result by a factor $2.042\,57$ even though $\epsilon\lesssim 0.1$.

Brown's Hamiltonian has been investigated by several authors in the past. In addition to the original papers by E.\ W.\ Brown, motivated by hierarchical triple star systems \citep{brown1,brown2,brown3}, explicit formulas for the Hamiltonian are given by \cite{sod75}, \cite{cuk04}, \cite{bv15}, \cite{luo16}, and others. However, the formulas given in these papers use different assumptions and different notation; curiously, even when these differences are resolved, the formulas often do not agree. One of the goals of this paper is to resolve these discrepancies. 

\section{Analysis}

\subsection{Orbit-averaged Hamiltonians}

We examine a hierarchical triple system consisting of an inner binary, with masses $m_0$ and $m_1$, and a third body of mass $m_2$ moving around the binary in a much larger orbit. 

Let $\bx_i$, $i=0,1,2$ be the positions of the three bodies in an inertial frame. We use Jacobi coordinates and momenta ($\br_i,\bp_i)$, $i=0,1,2$. The coordinate $\br_0$ is the position of the center of mass of the three bodies, $\br_1$ is the vector from $m_0$ to $m_1$, and $\br_2$ is the vector from the center of mass of $m_0$ and $m_1$ to $m_2$; thus
\begin{align}
\br_0&=\frac{m_0\bx_0+m_1\bx_1+m_2\bx_2}{m_0+m_1+m_2}, \nonumber \\
\br_1&=\bx_1-\bx_0, \nonumber \\
\br_2&=\bx_2-\frac{m_0\bx_0+m_1\bx_1}{m_0+m_1}.
\end{align}
The conjugate momenta are
\begin{align}
\bp_0&=m_0\dot\bx_0+m_1\dot\bx_1+m_2\dot\bx_2, \nonumber \\
\bp_1&=\frac{m_0m_1}{m_0+m_1}(\dot\bx_1-\dot\bx_0),\nonumber \\
\bp_2&=\frac{m_2}{m_0+m_1+m_2}[m_0(\dot\bx_2-\dot\bx_0)+m_1(\dot\bx_2-\bx_1)].
\end{align}
The Hamiltonian in Jacobi coordinates is
\begin{align}
H&=\frac{\bp_0^2}{2(m_0+m_1+m_2)}+\frac{\bp_1^2}{2\mu_1}+\frac{\bp_2^2}{2\mu_2}
-\frac{Gm_0m_1}{r_1}\nonumber \\&\quad -\frac{Gm_0m_2}{|\br_2+m_1\br_1/(m_0+m_1)|}-\frac{Gm_1m_2}{|\br_2-m_0\br_1/(m_0+m_1)|},
\end{align}
where $r_1=|\br_1|$ and the reduced masses are
\begin{equation}
\mu_1=\frac{m_0m_1}{m_0+m_1}, \quad
\mu_2=\frac{(m_0+m_1)m_2}{m_0+m_1+m_2}.
\label{eq:reduced}
\end{equation}
Since the Hamiltonian is independent of $\br_0$, the total momentum $\bp_0$ is conserved, so we can choose a frame in which it is zero.

We now expand the last two terms using the relation
\begin{equation}
\frac{1}{|\bR-\br|}=\sum_{l=0}^\infty
\frac{r^l}{R^{l+1}}P_l(\cos\gamma), \quad r<R,
\end{equation}
where $P_l(\cos\gamma)$ is a Legendre polynomial and $\cos\gamma=\br\cdot\bR/(rR)$ is the angle between the vectors $\br$ and $\bR$. The monopole, dipole, quadrupole, and octopole terms have $l=0,1,2,3$ respectively. Keeping terms up to $l=3$ we have
\begin{align}
&-\frac{Gm_0m_2}{|\br_2+m_1\br_1/(m_0+m_1)|}-\frac{Gm_1m_2}{|\br_2-m_0\br_1/(m_0+m_1)|}\nonumber \\
&= -\frac{Gm_2(m_0+m_1)}{r_2}-\frac{G m_0m_1m_2}{2(m_0+m_1)r_2^5}\big[3(\br_1\cdot\br_2)^2-r_1^2r_2^2\big]\nonumber \\
&\quad +\frac{Gm_0m_1m_2(m_1-m_0)}{2(m_0+m_1)^2r_2^7}\big[5(\br_1\cdot\br_2)^3-3r_1^2r_2^2(\br_1\cdot\br_2)\big].
\end{align}
We now rewrite the Hamiltonian as
\begin{equation}
H=H_\mathrm{K,in}+H_\mathrm{K,out} + H_\mathrm{quad} + H_\mathrm{oct}
\end{equation}
where the Kepler Hamiltonians are 
\begin{equation}
H_\mathrm{K,in}=\frac{\bp_1^2}{2\mu_1}-\frac{Gm_0m_1}{r_1}, \quad H_\mathrm{K,out}=\frac{\bp_2^2}{2\mu_2}-\frac{Gm_2(m_0+m_1)}{r_2},
\end{equation}
and the quadrupole and octopole Hamiltonians are
\begin{align}
H_\mathrm{quad}&=\frac{G
  m_0m_1m_2}{2(m_0+m_1)r_2^5}[r_1^2r_2^2-3(\br_1\cdot\br_2)^2], \\ H_\mathrm{oct}&=\frac{Gm_0m_1m_2(m_1-m_0)}{2(m_0+m_1)^2r_2^7}[(5(\br_1\cdot\br_2)^3-3r_1^2r_2^2(\br_1\cdot\br_2)].\nonumber
\end{align}

If we time-average the Hamiltonian over the inner orbit (``single-averaging'') we obtain
\begin{align}
\langle H_\mathrm{quad}\rangle_t &=\frac{Gm_0m_1m_2a_1^2}{4(m_0+m_1)r_2^5}[3(\bj_1\cdot\br_2)^2\!-15(\be_1\cdot\br_2)^2\!+(6e_1^2-1)r_2^2], \nonumber \\ \langle H_\mathrm{oct}\rangle_t &= \frac{5Gm_0m_1m_2(m_1-m_0)a_1^3}{16(m_0+m_1)^2r_2^7}\be_1\cdot\br_2\nonumber \\&\qquad\times [15(\bj_1\cdot\br_2)^2 - 35(\be_1\cdot\br_2)^2+(24e_1^2-3)r_2^2],
\label{eq:quads}
\end{align}
where $a_1$ is the semimajor axis of the inner binary, $\be_1$ is the eccentricity vector, which points from the origin at $m_0$ to the periapsis of $m_1$ and has magnitude $|\be_1|=e_1$, and $\bj_1$ is the dimensionless angular momentum, which is parallel to the angular momentum vector of the inner binary and has magnitude $|\bj_1|=(1-e_1^2)^{1/2}$. This and similar expressions below can be converted into orbital elements relative to a given $(x,y,z)$ coordinate system using the relations
\begin{align}
  e_x&=e\cos\Omega\cos\omega-e\cos I\sin\Omega\sin\omega, \nonumber \\
  e_y&= e\sin\Omega\cos\omega+e\cos I\cos\Omega\sin\omega,\nonumber \\
  e_z&=e\sin I\sin\omega, \nonumber\\
  j_x&=(1-e^2)^{1/2}\sin I\sin\Omega, \nonumber\\
j_y&=-(1-e^2)^{1/2}\sin I\cos\Omega, \nonumber\\ 
  j_z&=(1-e^2)^{1/2}\cos I,
  \label{eq:orbel}
\end{align}
where $\omega$ is the argument of periapsis, $\Omega$ is the longitude of the ascending node, and $I$ is the inclination.

In this paper we usually ignore changes in the orbit of the outer binary. This is a reasonable approximation whenever the angular momentum $L_2$ of the outer binary is much larger than the angular momentum $L_1$ of the inner binary; this criterion can be rewritten as
\begin{equation}
  \frac{L_2^2}{L_1^2}=
  \frac{m_2^2(m_0+m_1)^3}{m_0^2m_1^2(m_0+m_1+m_2)}\frac{a_2}{a_1}\frac{1-e_2^2}{1-e_1^2}
  \gg 1.
  \label{eq:jrat}
\end{equation}
Since $a_2\gg a_1$ in any hierarchical triple system, this inequality is usually satisfied unless $m_2\ll m_0+m_1$ or $1-e_2\ll 1$. For a treatment that does not make this assumption, see \cite{ford00}. 

Since the outer binary orbit is fixed, we adopt an inertial coordinate system in which the positive $x$-axis is parallel to $\be_2$ and the positive $z$-axis is parallel to $\bj_2$. Thus the outer orbital plane is the equatorial plane and the outer periapsis is on the $x$-axis. 

The equations of motion for the eccentricity vector $\be_1$ and the dimensionless angular-momentum vector $\bj_1$ are \citep{milank39,ttn09}
\begin{align}
\frac{d\bj_1}{dt}&=-\frac{1}{[G(m_0+m_1)a_1]^{1/2}}\left(\bj_1\times\frac{\p\langle H\rangle_t}{\p\bj_1} + \be_1\times\frac{\p\langle H\rangle_t}{\p\be_1}\right)\nonumber \\
\frac{d\be_1}{dt}&=-\frac{1}{[G(m_0+m_1)a_1]^{1/2}}\left(\be_1\times\frac{\p\langle H\rangle_t}{\p\bj_1} + \bj_1\times\frac{\p\langle H\rangle_t}{\p\be_1}\right).
\label{eq:milank}
    \end{align}
where $\langle H\rangle_t$ is the relevant orbit-averaged Hamiltonian and $a_1$, the semimajor axis of the inner binary, is invariant in the orbit-averaged approximation.  

The angle-action variables for the inner Kepler Hamiltonian $H_\mathrm{K,in}$ are $\ell_1,\omega_1,\Omega_1$ and their conjugate momenta (Delaunay variables) are
\begin{align}
\Lambda_1&=\frac{m_0m_1}{m_0+m_1}[G(m_0+m_1)a_1]^{1/2},\nonumber\\ 
L_1&=\Lambda_1(1-e_1^2)^{1/2},\nonumber\\  L_{z1}&=L_1\cos I_1.
\label{eq:delaunay}
\end{align}
   
If we also time-average the Hamiltonian over the outer orbit (``double-averaging''), we have
\begin{align}
\llangle H_\mathrm{quad}{\rrangle}_t &=
  \frac{G m_0m_1m_2a_1^2}{8(m_0+m_1)a_2^3(1-e_2^2)^{3/2}}\nonumber \\
  &\qquad \times \big[1-6e_1^2-3(\bj_1\cdot\jhat_2)^2+15(\be_1\cdot\jhat_2)^2\big]\label{eq:quadd}\\
  &= \frac{G m_0m_1m_2a_1^2}{8(m_0+m_1)a_2^3(1-e_2^2)^{3/2}}\nonumber\\
&\!\!\!\!\!\times\big(3\sin^2I_1-2-3e_1^2-3e_1^2\sin^2I_1+15e_1^2\sin^2I_1\sin^2\omega_1\big). \nonumber  
\end{align}
In these equations $a_2$ and $e_2$ are the semimajor axis and eccentricity of the outer binary. We have also introduced $\hat\be_2$ and $\jhat_2$, the unit vectors parallel to the eccentricity and angular-momentum vectors of the outer binary.

We shall also use the double-averaged octopole Hamiltonian
\begin{align}
\llangle H_\mathrm{oct}{\rrangle}_t &=
\frac{15Gm_0m_1m_2(m_0-m_1)a_1^3}{64(m_0+m_1)^2a_2^4(1-e_2^2)^{5/2}}e_2\nonumber \\
&\quad\times \Big\{\be_1\cdot\hat\be_2\big[8e_1^2-1+5(\bj_1\cdot\jhat_2)^2-35(\be_1\cdot\jhat_2)^2\big] \nonumber \\
&\qquad +10(\be_1\cdot\jhat_2)(\bj_1\cdot\hat\be_2)(\bj_1\cdot\jhat_2)\Big\}\nonumber \\
  &=\frac{15Gm_0m_1m_2(m_0-m_1)a_1^3}{64(m_0+m_1)^2a_2^4(1-e_2^2)^{5/2}}e_1e_2\nonumber \\
  &\quad\times \Big\{(\cos\Omega_1\cos\omega_1-\cos I_1\sin\Omega_1\sin\omega_1)\big[8e_1^2-1\nonumber \\
  &\qquad\quad +5(1-e_1^2)\cos^2I_1-35e_1^2\sin^2I_1\sin^2\omega_1\big] \nonumber \\
&\qquad\quad + 10(1-e_1^2)\sin^2I_1\cos I_1\sin\omega_1\sin\Omega_1\Big\}. 
\label{eq:fordo}
\end{align}
Notice that the octopole Hamiltonian vanishes when the outer binary is on a circular orbit or when the masses of the two inner components are equal. 

\subsection{Brown's Hamiltonian}

\label{sec:q2}

Our goal is to evaluate the long-term effects of the single-averaged quadrupole Hamiltonian $\langle H_\mathrm{quad}\rangle_t$ (eq.\ \ref{eq:quads}) to higher order in the strength $\epsilon$ of the perturbation from the distant body. We approach this problem using Poincar\'e--von Zeipel perturbation theory, in which we carry out a canonical transformation to new coordinates and momenta that eliminates secular terms in the equations of motion. 

Since we have already averaged over the orbital motion of the inner binary, its semimajor axis is fixed and we can restrict ourselves to a phase space with two degrees of freedom, having coordinates $\omega_1$ and $\Omega_1$ and conjugate momenta $L_1$ and $L_{z1}$.  

To carry out this analysis as generally as possible, it proves useful to introduce a fictitious time $\tau$ that varies from 0 to $2\pi$ over one orbital period of the outer binary. Thus $\tau=\ell_2+g(\ell_2)$ where $\ell_2$ is the mean anomaly of the outer binary and $g(\ell_2)$ is an odd periodic function with period $2\pi$ and $g'(\ell_2)>-1$. The actions and angles satisfy Hamilton's equations in the fictitious time if the single-averaged quadrupole Hamiltonian $\langle H_\mathrm{quad}\rangle_t$ is replaced by
\begin{equation}
\epsilon^2H_\tau(\bq,\bp,\tau)\equiv\epsilon^2w(\tau)\langle H_\mathrm{quad}\rangle_t\ \mbox{where}\ w(\tau)\equiv \frac{1}{n_2[1+g'(\ell_2)]}.
\end{equation}
Here $\epsilon$ is a ordering parameter defined by equations (\ref{eq:epsdef1}) or (\ref{eq:epsdef2}), which will be set to unity later, $\langle H_\mathrm{quad}\rangle_t$ depends on time $t$ through the time dependence of the position of the outer body, $\br_2(t)$, and  $n_2=[G(m_0+m_1+m_2)/a_2^3]^{1/2}$ is the mean motion of the outer binary. 

Note that the time average $\langle X\rangle_t$ and the $\tau$ average $\langle w(\tau) X\rangle_\tau$ for any quantity $X$ are related by
\begin{equation}
\langle w(\tau)X\rangle_\tau =\frac{1}{n_2}\langle X \rangle_t.
\label{eq:wdef}
\end{equation}

Let $S(\bq,\bp',\tau)=\bq\cdot\bp'+\epsilon s(\bq,\bp',\tau)$ be the generating function for a canonical transformation to new variables $(\bq',\bp')$. Then
\begin{equation}
\bq'=\frac{\p S}{\p \bp'}=\bq+\epsilon\frac{\p s}{\p\bp}, \quad \bp =\frac{\p S}{\p \bq}=\bp'+\epsilon\frac{\p s}{\p\bq}.
\end{equation}
The transformed Hamiltonian is 
\begin{align}
\epsilon^2 H'_\tau(\bq',\bp',\tau)&=\epsilon^2 H_\tau(\bq,\bp,\tau)+\epsilon^2 \frac{\p s}{\p\tau}(\bq,\bp',\tau) \nonumber \\
&=\epsilon^2 H_\tau + \epsilon^2\frac{\p s}{\p \tau}
+\epsilon^3\sum_i\bigg(-\frac{\p H_\tau}{\p q'_i}\frac{\p s}{\p p'_i} \nonumber \\&\qquad + \frac{\p
  H_\tau}{\p p'_i}\frac{\p s}{\p q'_i} - \frac{\p^2s}{\p q'_i \p
  \tau}\frac{\p s}{\p p'_i}\bigg) +\mbox{O}(\epsilon^4), 
\end{align}
where all functions in the last equation are evaluated at the primed coordinates and momenta. Notice that the term $\p s/\p\tau$ is of order $\epsilon^2$ because $s$ varies with the orbital frequency of the outer binary, which is of order $\epsilon$. 

We denote the average of $H_\tau$ over the fictitious time at fixed phase-space coordinates by $\langle H_\tau\rangle_\tau$, and we write $\{H_\tau\}_\tau=H_\tau-\langle H_\tau\rangle_\tau$. We now set 
\begin{equation}
\frac{\p s}{\p \tau}=-\{H_\tau\}_\tau.
\label{eq:sdef}
\end{equation}
This relation can be integrated to determine the generating function $s(\bq,\bp',\tau)$, and the constant of integration can be chosen so that $\langle s\rangle_\tau=0$. Thus the differences between the primed and unprimed coordinates and momenta also oscillate. Then
\begin{align}
\epsilon^2 H'_\tau(\bq',\bp',\tau)&=\epsilon^2 \langle H_\tau\rangle_\tau  \nonumber \\
&\quad +\epsilon^3\sum_i\left(-\frac{\p \langle H_\tau\rangle_\tau}{\p q'_i}\frac{\p s}{\p p'_i} + \frac{\p H_\tau}{\p p'_i}\frac{\p s}{\p q'_i} \right) +\mbox{O}(\epsilon^4). 
\label{eq:brown1}
\end{align}
We may now average the Hamiltonian over fictitious time to obtain
\begin{equation}
\epsilon \langle H_\tau'(\bq',\bp',t)\rangle_\tau=\epsilon^2 \langle H_\tau\rangle_\tau 
+\epsilon^3\sum_i\left\langle\frac{\p H_\tau}{\p p'_i}\frac{\p s}{\p q'_i}\right\rangle_\tau +\mbox{O}(\epsilon^4). 
\end{equation}
In this equation the first term in brackets in equation (\ref{eq:brown1}) has vanished upon averaging because the first factor is constant and the second averages to zero. 

We now make several changes in notation. We drop all terms that are $\mbox{O}(\epsilon^4)$ or higher. We drop the primes on the coordinates and momenta; this is acceptable because the primed and unprimed variables only differ by fluctuating terms that are $\mbox{O}(\epsilon)$. We also set $\epsilon=1$. Moreover
\begin{align}
    \langle H_\tau\rangle_\tau&=\frac{1}{2\pi}\!\int_0^{2\pi}\!w(\tau)d\tau\,\langle H_\mathrm{quad}\rangle_t \nonumber \\ &=\frac{1}{2\pi}\!\int_0^P \!dt\langle H_\mathrm{quad}\rangle_t =\frac{1}{n_2}\llangle H_\mathrm{quad}\rrangle_t,
\end{align}
the double-averaged Hamiltonian of equation (\ref{eq:quadd}).  Thus our final approximation to the quadrupole Hamiltonian is
\begin{equation}
\frac{1}{n_2}\left(\llangle H_\mathrm{quad} {\rrangle}_t + H_\mathrm{B}\right),
\end{equation}
where Brown's Hamiltonian is
\begin{align}
    H_\mathrm{B}&\equiv n_2\left\langle w(\tau)\frac{\p \langle H_\mathrm{quad}\rangle_t}{\p L_1}\frac{\p s}{\p \omega_1} + w(\tau)\frac{\p \langle H_\mathrm{quad}\rangle_t}{\p L_{z1}}\frac{\p s}{\p \Omega_1}\right\rangle_\tau \nonumber \\
    &=\frac{n_2}{2\pi}\int_0^{2\pi}\!\!w(\tau)d\tau \left(\frac{\p \langle H_\mathrm{quad}\rangle_t}{\p L_1}\frac{\p s}{\p \omega_1} + \frac{\p \langle H_\mathrm{quad}\rangle_t}{\p L_{z1}}\frac{\p
    s}{\p \Omega_1}\right).  
\label{eq:hqqav}
\end{align}

The single-averaged quadrupole potential (\ref{eq:quads}) can be written 
\begin{align}
\langle H_\mathrm{quad}\rangle_t &= \frac{Gm_0m_1m_2a_1^2}{8(m_0+m_1)a_2^3(1- e_2^2)^3} \big[\alpha(\be_1,\bj_1)F_\alpha(e_2,\tau) \nonumber \\&\quad +\beta(\be_1,\bj_1)F_\beta(e_2,\tau) + \gamma(\be_1,\bj_1)F_\gamma(e_2,\tau)\big],
\end{align}
where
\begin{align}
F_\alpha(e,\tau)&=(1+e\cos f_2)^3\cos 2f_2, \nonumber \\ F_\beta(e,\tau)&=(1+e\cos f_2)^3\sin2f_2, \nonumber\\ F_\gamma(e,\tau)&=(1+e\cos f_2)^3;
\label{eq:fdef}
\end{align}
in these equations the true anomaly $f_2$ of the outer binary is considered to be a function of the fictitious time $\tau$. We also have 
\begin{align}
\alpha(\be,\bj)&=3j_x^2-3j_y^2-15e_x^2+15e_y^2, \nonumber \\ \beta(\be,\bj)&=6j_xj_y-30e_xe_y, \nonumber \\ \gamma(\be,\bj)&=12e^2-2+3j_x^2+3j_y^2-15e_x^2-15e_y^2\nonumber \\ &=1-6e^2-3j_z^2+15e_z^2.
\end{align}

It is straightforward to show that 
\begin{equation}
\langle F_\alpha\rangle_t =\langle F_\beta\rangle_t=0, \quad \langle F_\gamma\rangle_t = (1-e_2^2)^{3/2}.
\label{eq:fav}
\end{equation}
Therefore the double-averaged quadrupole Hamiltonian is 
\begin{align}
\llangle H_\mathrm{quad}{\rrangle}_t &= \frac{Gm_0m_1m_2a_1^2}{8(m_0+m_1)a_2^3(1-e_2^2)^3}\gamma(\be_1,\bj_1)\langle F_\gamma(e_2,f_2)\rangle_t \nonumber \\ &=\frac{Gm_0m_1m_2a_1^2}{8(m_0+m_1)a_2^3(1-e_2^2)^{3/2}}\gamma(\be_1,\bj_1),
\label{eq:hquadd}
\end{align}
consistent with equation (\ref{eq:quadd}). 

The fluctuating part of the single-averaged Hamiltonian is 
\begin{align}
\{H_\tau\}_\tau & = H_\tau - \langle H_\tau\rangle_\tau =w(\tau)\langle H_\mathrm{quad}\rangle_t -
\frac{1}{n_2}\llangle H_\mathrm{quad}{\rrangle}_t \nonumber \\ &=\frac{Gm_0m_1m_2a_1^2}{8(m_0+m_1)a_2^3(1-e_2^2)^3}\big[\alpha(\be_1,\bj_1)w(\tau)F_\alpha(e_2,f_2) \nonumber \\ &\quad + \beta(\be_1,\bj_1)w(\tau)F_\beta(e_2,f_2) + \gamma(\be_1,\bj_1)w(\tau)F_\gamma(e_2,f_2)\nonumber \\ &\qquad-\gamma(\be_1,\bj_1)(1-e_2^2)^{3/2}/n_2\big].
\end{align}

We define
\begin{align}
G_\alpha(e,\tau)&\equiv \int_0^\tau w(\tau')d\tau'\,F_\alpha(e,\tau'),\nonumber \\G_\beta(e,\tau)&\equiv \int_0^\tau w(\tau')d\tau'\,F_\beta(e,\tau'),\nonumber \\G_\gamma(e,\tau)&\equiv \int_0^\tau w(\tau')d\tau'\,F_\gamma(e,\tau')-(1-e^2)^{3/2}\tau/n_2.  
\label{eq:gdef}
\end{align}
Notice that
\begin{align}
0&=\langle F_\alpha\rangle_t=\frac{n_2}{2\pi} \int_0^{2\pi}\!\!\!w(\tau)d\tau\,F_\alpha(e,\tau)=\frac{n_2}{2\pi}\int_0^{2\pi}\!\!\!d\tau\, dG_\alpha(e,\tau)/d\tau\nonumber \\
&=\frac{n_2}{2\pi}[G_\alpha(2\pi)-G_\alpha(0)].
\end{align}
Thus $G_\alpha(e,\tau)$ is periodic; the same is true for $G_\beta(e,\tau)$ and $G_\gamma(e,\tau)$. We define the fluctuating part of $G_\alpha(e,\tau)$ to be $\{G_\alpha\}_\tau\equiv G_\alpha - \langle G_\alpha\rangle_\tau$ with similar definitions for $\{ G_\beta\}_\tau$ and $\{ G_\gamma\}_\tau$. Since $\tau$ is an odd function of the anomaly $f$, $G_\alpha(e,\tau)$ is odd so $\langle G_\alpha\rangle_\tau=0$ and $\{G_\alpha\}_\tau= G_\alpha$. Similarly $\{G_\gamma\}_\tau= G_\gamma$.

 We may now integrate equation (\ref{eq:sdef}) to find the canonical transformation,
\begin{align}
  s&= -\frac{Gm_0m_1m_2a_1^2}{8(m_0+m_1)a_2^3(1-e_2^2)^3}\Big[\alpha(\be_1,\bj_1)\{G_\alpha(e_2,\tau)\}_\tau \nonumber \\&\qquad + \beta(\be_1,\bj_1)\{G_\beta(e_2,\tau)\}_\tau+\gamma(\be_1,\bj_1)\{G_\gamma(e_2,\tau)\}_\tau\Big].
\end{align}
We use only the fluctuating parts of $G_{\{\alpha,\beta,\gamma\}}(e_2,\tau)$ to ensure that $\langle s\rangle_\tau=0$.

To evaluate the orbit average in equation (\ref{eq:hqqav}) we need orbit averages of products such as $w(\tau)F_\alpha \{G_\alpha\}_\tau$, $w(\tau)F_\alpha \{G_\beta\}_\tau$, etc. We have assumed that the fictitious time is an odd function of the mean anomaly, as is the true anomaly. Therefore $w(\tau)$, $F_\alpha(e,\tau)$ and $F_\gamma(e,\tau)$ are even functions of the fictitious time, and $F_\beta(e,\tau)$ is odd. Similarly $\{G_\alpha(e,\tau)\}_\tau$ and $\{G_\gamma(e,\tau)\}_\tau$ are odd and $\{G_\beta(e,\tau)\}_\tau$ is even. 
Therefore there are only four non-zero
products. Moreover, 
\begin{align}
&\langle w(\tau)F_\alpha \{G_\beta\}_\tau\rangle_\tau=\left\langle \frac{d\{G_\alpha\}_\tau}{d\tau} \{G_\beta\}_\tau\right\rangle_\tau\nonumber \\
&=\frac{1}{2\pi}\{G_\alpha(\tau)\}_\tau\{G_\beta(\tau)\}_\tau|_{0}^{2\pi}-\left\langle \{G_\alpha \}_\tau\frac{d\{G_\beta\}}{d\tau}\right\rangle_\tau\nonumber \\&=
-\langle w(\tau)F_\beta \{G_\alpha\}_\tau\rangle_\tau.
\end{align}
Similarly $\langle w(\tau) F_\beta \{G_\gamma\}_\tau\rangle_\tau=-\langle w(\tau) F_\gamma \{G_\beta\}_\tau\rangle_\tau$. Thus the only relations we need are encapsulated in two new functions:
\begin{align}
A(e_2)&\equiv\frac{n_2^2}{(1-e_2^2)^6}\langle w(\tau)F_\alpha \{G_\beta\}_\tau \rangle_\tau \nonumber \\&= -\frac{n_2^2}{(1-e_2^2)^6}\langle w(\tau) F_\beta 
\{G_\alpha\}_\tau \rangle_\tau , \nonumber \\
C(e_2)&\equiv \frac{n_2^2}{(1-e_2^2)^6}\langle w(\tau)F_\beta \{G_\gamma\}_\tau\rangle_\tau \nonumber \\&=-\frac{n_2^2}{(1-e_2^2)^6}\langle w(\tau)F_\gamma \{G_\beta\}_\tau\rangle_\tau .
\label{eq:abdef}
\end{align}

Then equation (\ref{eq:hqqav}) gives 
\begin{align}
H_\mathrm{B}&=\frac{G^{3/2}m_0^2m_1^2m_2^2a_1^4}{64(m_0+m_1)^2(m_0+m_1+m_2)^{1/2}a_2^{9/2}}\nonumber\\&\qquad \times 
\big[A(e_2)\{\alpha,\beta\}_\mathrm{P}+C(e_2)\{\beta,\gamma\}_\mathrm{P}\big]
\label{eq:hqqdef}
\end{align}
Here 
\begin{align}
    \{\alpha,\beta\}_\mathrm{P}&\equiv \sum_i\left(\frac{\p\alpha}{\p q_i}\frac{\p\beta}{\p p_i}-\frac{\p\beta}{\p q_i}\frac{\p\alpha}{\p p_i}\right)\nonumber \\
    &=\frac{\p\alpha}{\p \omega_1}\frac{\p\beta}{\p L_1}+\frac{\p\alpha}{\p \Omega}\frac{\p\beta}{\p L_{z1}}-\frac{\p\beta}{\p \omega_1}\frac{\p\alpha}{\p L_1}-\frac{\p\beta}{\p \Omega}\frac{\p\alpha}{\p L_{z1}}
\end{align}
is a Poisson bracket.

The Poisson brackets can be evaluated most easily using the relations
\begin{align}
\{ j_i,j_j\}_\mathrm{P}&=\frac{1}{\Lambda_1}\sum_{k=1}^3\epsilon_{ijk}j_k, \quad \{
e_i,e_j\}_\mathrm{P}=\frac{1}{\Lambda_1}\sum_{k=1}^3\epsilon_{ijk}j_k, \nonumber \\ \{
e_i,j_j\}_\mathrm{P}&=\{ j_i,e_j\}_\mathrm{P}=\frac{1}{\Lambda_1}\sum_{k=1}^3\epsilon_{ijk}e_k,
\end{align}
where $\epsilon_{ijk}$ is the antisymmetric tensor and $\Lambda_1$ is
given by equation (\ref{eq:delaunay}).

The Poisson brackets are
\begin{align}
\{\alpha,\beta\}_\mathrm{P} &=\frac{36(m_0+m_1)^{1/2}}{m_0m_1(Ga_1)^{1/2}} (\bj_1\cdot\jhat_2)\big[24e_1^2-15(\be_1\cdot\jhat_2)^2 \nonumber\\[2pt]&\qquad- (\bj_1\cdot\jhat_2)^2+1\big],\nonumber \\[2pt] \{\beta,\gamma\}_\mathrm{P} &=\frac{36(m_0+m_1)^{1/2}}{m_0m_1(Ga_1)^{1/2}}\big\{(\bj_1\cdot\jhat_2)[1-2(\bj_1\cdot\hat\be_2)^2-(\bj_1\cdot\jhat_2)^2\nonumber \\[2pt] &\qquad +4e_1^2-10(\be_1\cdot\hat\be_2)^2-15(\be_1\cdot\jhat_2)^2] \nonumber \\[2pt]&\qquad -20(\be_1\cdot\hat\be_2)(\bj_1\cdot\hat\be_2)(\be_1\cdot\jhat_2)\big\}.
\end{align}
where $\jhat_2$ and $\hat\be_2$ are unit vectors parallel to the angular-momentum and eccentricity vector of the outer orbit. In terms of the orbital elements,
\begin{align}
\{\alpha,\beta\}_\mathrm{P} &=\frac{36(m_0+m_1)^{1/2}\cos  I_1(1-e_1^2)^{1/2}}{m_0m_1(Ga_1)^{1/2}}\nonumber\\[2pt]&\times\big[1+24e_1^2-(1-e_1^2)\cos^2I_1 - 15e_1^2\sin^2I_1\sin^2\omega_1\big], \nonumber \\[2pt]
\{\beta,\gamma\}_\mathrm{P} &=\frac{9(m_0+m_1)^{1/2}(1-e_1^2)^{1/2}}{m_0m_1(Ga_1)^{1/2}}\nonumber\\[2pt]&\times \big\{2\cos I_1\sin^2I_1(2-17e_1^2)\cos2\Omega_1  \nonumber\\[2pt]
&-5e_1^2[(1-\cos I_1)^2(3\cos I_1+2)\cos 2(\omega_1-\Omega_1) \nonumber\\[2pt]&+ (1+\cos I_1)^2(3\cos I_1-2)\cos2(\omega_1+\Omega_1)]\big\}.
\label{eq:aborbel}
\end{align}

The Hamiltonian (\ref{eq:hqqdef}) can now be written  
\begin{align}
H_\mathrm{B}&=
\frac{9Gm_0m_1m_2^2a_1^{7/2}}{16(m_0+m_1)^{3/2}(m_0+m_1+m_2)^{1/2}a_2^{9/2}}\nonumber \\&\ \times \Big\llbracket A(e_2) (\bj_1\cdot\jhat_2)\big[24e_1^2-15(\be_1\cdot\jhat_2)^2-(\bj_1\cdot\jhat_2)^2+1\big]\label{eq:hbrown}\\
&\ +C(e_2)\big\{(\bj_1\cdot\jhat_2)[1-2(\bj_1\cdot\hat\be_2)^2-(\bj_1\cdot\jhat_2)^2+4e_1^2 \nonumber \\ &\ -10(\be_1\cdot\hat\be_2)^2-15(\be_1\cdot\jhat_2)^2]-20(\be_1\cdot\hat\be_2)(\bj_1\cdot\hat\be_2)(\be_1\cdot\jhat_2)\big\}\Big\rrbracket.\nonumber 
  \end{align}
This can be rewritten in terms of orbital elements using equations (\ref{eq:orbel}) or (\ref{eq:aborbel}). 

Brown's Hamiltonian represents the nonlinear effects of the quadrupole potential from the outer body -- the Hamiltonian is proportional to the square of the perturbing mass $m_2$ in the limit when $m_2 \ll m_0,m_1$. Generalized (but very complicated) analogs of Brown's Hamiltonian for octopole and higher order potentials are given by \cite{lei18}.

Brown's Hamiltonian $H_\mathrm{B}$, together with the double-averaged quadrupole and octopole Hamiltonians $\llangle H_\mathrm{quad}{\rrangle}_t$ (eq.\ \ref{eq:quadd}) and $\llangle H_\mathrm{oct}{\rrangle}_t$ (eq.\ \ref{eq:fordo}) and the Milankovich equations of motion (\ref{eq:milank}), determine the long-term evolution of the eccentricity and angular-momentum vectors of the inner binary if variations in the orbit of the outer binary can be neglected. The fictitious time differs from the real time, but only by an amount that oscillates with period $2\pi/n_2$, so in calculating the long-term evolution of the inner binary we can replace the fictitious time by the real time. 

We now examine the functions $A(e_2)$ and $C(e_2)$ in the Hamiltonian, which encapsulate the dependence of Brown's Hamiltonian on the eccentricity $e_2$ of the outer binary.

First consider the definitions of $G_{\{\alpha,\beta,\gamma\}}(e,\tau)$ in equations (\ref{eq:gdef}). Since $w(\tau)d\tau=dt$, the first of these can be written 
\begin{align}
    G_\alpha(e,\tau)&=\frac{(1-e^2)^{3/2}}{n_2}\int_0^\tau \frac{df F_\alpha}{(1+e\cos f)^2}\nonumber \\&= \frac{(1-e^2)^{3/2}}{n_2}\int_0^\tau \df(1+e\cos f)\cos 2f\nonumber \\&= \frac{(1-e^2)^{3/2}}{3n_2}\sin f(3\cos f+ 2e+e\cos 2f),
\end{align}
in which $n_2=[G(m_0+m_1+m_2)/a_2^3]^{1/2}$ is the mean motion of the outer binary and the true anomaly $f$ is a function of the fictitious time $\tau$. 

Similarly,
\begin{align}
  G_\beta(e,f)&=\frac{(1-e^2)^{3/2}}{6n_2}(3+4e-3e\cos f-3\cos 2f-e\cos 3f), \nonumber \\
 G_\gamma(e,f)&=\frac{(1-e^2)^{3/2}}{n_2}(f+e\sin f-\tau).
 \label{eq:gbetadef}
\end{align}

From the first of equations (\ref{eq:abdef}),
\begin{equation}
    A(e_2)=\frac{n_2^2}{(1-e_2^2)^6}\left[\langle w(\tau)F_\alpha G_\beta\rangle_\tau - \langle w(\tau)F_\alpha\rangle_\tau\langle G_\beta\rangle_\tau\right].
\end{equation}
Now $\langle w(\tau)F_\alpha\rangle_\tau=\langle F_\alpha\rangle_t/n_2$ (eq.\ \ref{eq:wdef}) and from the first of equations (\ref{eq:fav}) this vanishes. Thus
\begin{equation}
    A(e_2)=\frac{n_2}{(1-e_2^2)^6}\langle F_\alpha G_\beta\rangle_t;
\end{equation}
this time average is straightforward to evaluate using equations (\ref{eq:fdef}) and (\ref{eq:gbetadef}), and we find
\begin{equation}
A(e_2)=-\frac{(3+2e_2^2)}{12(1-e_2^2)^3},
\end{equation}
a result that is independent of the choice of the fictitious time $\tau$.

Similarly,
\begin{equation}
    C(e_2)=\frac{n_2^2}{(1-e_2^2)^6}\left[\langle w(\tau)F_\beta G_\gamma\rangle_\tau - \langle w(\tau)F_\beta\rangle_\tau\langle G_\gamma\rangle_\tau\right];
\end{equation}
and $\langle w(\tau)F_\beta\rangle_\tau=\langle F_\beta\rangle_t/n_2$ which vanishes from the second of equations (\ref{eq:fav}).
Thus
\begin{align}
    C(e_2)&=\frac{1}{(1-e_2^2)^{9/2}}\langle (1+e_2\cos f)^3\sin 2f (f+e_2\sin f -\tau)\rangle_t\nonumber \\
    &=\frac{1}{(1-e_2^2)^3}\int_0^{2\pi}\frac{df}{2\pi}(1+e_2\cos f)\sin 2f (f+e_2\sin f -\tau). 
    \label{eq:bdef}
\end{align}

In contrast to $A(e_2)$, the function $C(e_2)$ depends on the definition of the fictitious time $\tau$. Since $\tau$ is an odd function of the mean anomaly or true anomaly, we can write
\begin{equation}
\tau=f+\sum_{j=1}^\infty a_j\sin jf.
\end{equation}
Substituting this expansion in equation (\ref{eq:bdef}) and evaluating the integral, we find 
\begin{equation}
C(e_2)=\frac{1}{(1-e_2^2)^3}\left[\tfrac{1}{4}e_2(e_2-a_1)-\half a_2 -\tfrac{1}{4}e_2a_3\right].
\end{equation}
The coefficients $a_4$, $a_5$, etc.\  have no effect; thus $C(e_2)$ depends only on the three parameters $a_1, a_2, a_3.$

Rather than parametrize $C(e_2)$ by these coefficients, we describe three physically motivated choices for the fictitious time which span the range of possible behavior we can expect.

\paragraph*{1. Fictitious time = mean anomaly} The traditional choice is to set $\tau=\ell=n_2t$ where $\ell$ is the mean anomaly and $n_2$ is the mean motion of the outer binary. Thus the fictitious and true time are the same except for a proportionality constant. 

The standard Fourier expansion of the mean anomaly in terms of the true anomaly yields \citep[e.g.,][p.\ 65]{bc61} 
\begin{align}
a_1&=-2e_2, \nonumber\\ a_2&=\frac{2}{e_2^2}[1-(1-e_2^2)^{1/2}]^2[\half+(1-e_2^2)^{1/2}], \nonumber\\ a_3&=-\frac{2}{e_2^3}[1-(1-e_2^2)^{1/2}]^3[\tfrac{1}{3}+(1-e_2^2)^{1/2}],
\end{align}
and 
\begin{equation}
C(e_2)= 
\frac{1}{12e_2^2(1-e_2^2)^3}[4(1-e_2^2)^{3/2}-4+6e_2^2+3e_2^4].
\label{eq:abmean}
\end{equation}
The function $C(e_2)$ is defined as $e_2\to 0$ by its limit,
$C(e)=\frac{3}{8}e^2 + \mbox{O}(e^4)$. 

When equation (\ref{eq:hbrown}) is evaluated with this expression for $C(e_2)$ we call the Hamiltonian $H_\mathrm{B1}$.

\paragraph*{2. Fictitious time = true anomaly} In this case 
$a_1=a_2=a_3=0$, and 
\begin{equation}
C(e_2)=\frac{e_2^2}{4(1-e_2^2)^3}.
\label{eq:abtrue}
\end{equation}

With this expression for $C(e_2)$ we call the Hamiltonian $H_\mathrm{B2}$. 

\paragraph*{3. Fictitious time = $\mathbf{f+e\,\mbox{sin\,} f}$}
By choosing $\tau=f+e\sin f$ we have 
$a_1=e$, $a_2=0$, $a_3=0$, so 
\begin{equation}
C(e_2)=0.
\label{eq:abnew}
\end{equation}

With this expression for $C(e_2)$ we call the Hamiltonian $H_\mathrm{B3}$. 

\bigskip

These examples show that $C(e_2)$ depends on the choice of the relation between the fictitious time and the true time. Thus, if the orbit of the outer binary is eccentric, the Hamiltonian $H_\mathrm{B}$ and hence the equations of motion will depend on the choice of fictitious time. However, this dependence yields solutions of the equations of motion that differ only by terms that oscillate and hence the long-term solutions of the equations of motion are the same for any fictitious time. The dependence of Brown's Hamiltonian on the choice of fictitious time can be thought of as a kind of gauge freedom in the definition of the Hamiltonian. 

Given this freedom, by far the simplest expression for Brown's Hamiltonian is obtained with choice 3 for the fictitious time. In this case we have 
\begin{align}
H_\mathrm{B}&=H_\mathrm{B3}=
-\frac{3Gm_0m_1m_2^2a_1^{7/2}(3+2e_2^2)}{64(m_0+m_1)^{3/2}(m_0+m_1+m_2)^{1/2}a_2^{9/2}(1-e_2^2)^3}\nonumber \\&\ \times (\bj_1\cdot\jhat_2)\big[24e_1^2-15(\be_1\cdot\jhat_2)^2-(\bj_1\cdot\jhat_2)^2+1\big]\nonumber\\
&\qquad\ \  =
-\frac{3Gm_0m_1m_2^2a_1^{7/2}(3+2e_2^2)\cos I_1(1-e_1^2)^{1/2}}{64(m_0+m_1)^{3/2}(m_0+m_1+m_2)^{1/2}a_2^{9/2}(1-e_2^2)^3}\nonumber \\&\ \times[1+24e_1^2-(1-e_1^2)\cos^2I_1-15 e_1^2\sin^2I_1\sin^2\omega_1].\label{eq:hb3}
  \end{align}

Notice that the Hamiltonian $H_\mathrm{B3}$ is independent of the longitude of node $\Omega_1$ so the conjugate momentum $L_{z1}$ is conserved, as is $\bj_1\cdot\jhat_2$. Thus both the double-averaged quadrupole Hamiltonian $\llangle H_\mathrm{quad}\rrangle_t$ and Brown's Hamiltonian $H_\mathrm{B3}$ conserve $L_{z1}$, so the motion under the influence of the sum of these Hamiltonians is integrable. Even when $H_\mathrm{B3}$ is much larger than the octopole Hamiltonian $\llangle H_\mathrm{oct}\rrangle_t$, it is important to include the latter because it enables variations in $L_{z1}$ which allow the system to explore a much larger volume of phase space.

\section{Comparison to other work}

\subsection{Brown (1936a,b,c)}

Most of this analysis is contained in three remarkable papers by E.\ W.\ Brown (1936a,b,c). 

The double-averaged octopole Hamiltonian (\ref{eq:fordo}) is equivalent to the equation starting ``$k_1k_2(\frac{5}{16}e+\cdots$'' on p.\ 64 of \cite{brown2}, except that (i) Brown's expression is $-\llangle H_\mathrm{oct}{\rrangle}_t$; (ii) Brown's expression is per unit mass, so must be multiplied by the reduced mass of the inner binary, $m_0m_1/(m_0+m_1)$ (eq.\ \ref{eq:reduced}); (iii) in Brown's next equation $k_3$ should be replaced by $k_2$. This paper also introduced the concept of using the true anomaly as the fictitious time (``it is mainly due to this change that the problem and its solution can be so greatly simplified''). 

Brown's Hamiltonian is discussed in \cite{brown3}. In that paper the relevant Hamiltonian is written on p.\ 119 as $R'=R_c+\half(R_f,S)_c$ (the bracket is Brown's notation for minus the Poisson bracket). To compare this to our results we must (i) specialize to the case in which the masses satisfy the hierarchy $m_1\ll m_0\ll m_2$; (ii) recognize that $R'$ is \emph{minus} the Hamiltonian; (iii) drop an unimportant constant term; (iv) multiply $R'$ by $n_2$, the mean motion of the outer binary, since $R'$ is the Hamiltonian for the fictitious time $\tau=f_2$ rather than the true time. With these changes, Brown's expression for $R_c$ on p.\ 121 is equivalent to our equation (\ref{eq:quadd}) for the double-averaged quadrupole Hamiltonian.

Brown evaluates the second term in the expression for $R'$, $\half(R_f,S)_c$, in the special case where the orbit of the outer binary is circular ($e_2=0$). In this case our expression for Brown's Hamiltonian, equation (\ref{eq:hbrown}) or (\ref{eq:hb3}), simplifies to  
\begin{align}
\frac{H_\mathrm{B}}{m_1}&=-
\frac{9Gm_2^{3/2}a_1^{7/2}}{64m_0^{1/2}a_2^{9/2}}(\bj_1\cdot\jhat_2)\big[24e_1^2-15(\be_1\cdot\jhat_2)^2-(\bj_1\cdot\jhat_2)^2+1\big]\nonumber \\&=- \frac{9n_2^3a_1^2\cos I_1(1-e_1^2)^{1/2}}{128n_1} \label{eq:cuk} \\
&\quad\times [2+33e_1^2+(17e_1^2-2)\cos^2I_1+15e_1^2\sin^2I_1\cos2\omega_1]\nonumber
  \end{align}
where $n_1=(Gm_0/a_1^3)^{1/2}$ is the mean motion of the inner binary. This is equivalent to the expression given by Brown for $\half(R_f,S)_c$ on p.\ 121, as was recognized by \cite{bv15}. 

Furthermore Brown shows on p.\ 120 that when the outer eccentri\-city $e_2$ is non-zero, Brown's Hamiltonian can be obtained from equation (\ref{eq:cuk}) simply by multiplying that expression by $(1+\frac{2}{3}e_2^2)/(1-e_2^2)^3$, a result that is equivalent to our expression $H_\mathrm{B3}$ when $m_1\ll m_0\ll m_2$ (cf.\ eq.\ \ref{eq:hb3}). 

Therefore almost all of the important results in this paper and the recent literature are contained in Brown's 1936 papers. 

\subsection{S\"oderhjelm (1975)}

S\"oderhjelm's equation (14) gives the sum of the double-averaged quadrupole Hamiltonian (\ref{eq:quadd}) -- in his notation, the term proportional to $\half k_1L_1$ -- and Brown's Hamiltonian $H_\mathrm{B3}$ -- the term proportional to $\half k_1k_2L_1$. To compare this to our results\footnote{Note that S\"oderhjelm and \cite{will21} work in a coordinate system in which the positive $z$-axis is parallel to the total angular momentum of the system, whereas we assume that the $z$-axis is parallel to the angular momentum of the outer binary and the positive $x$-axis is parallel to the eccentricity vector of the outer binary. In the limit we have assumed, in which the angular momentum of the outer binary is much larger than that of the inner binary, the two coordinate systems are equivalent, if the orbital elements are $I_2=0$, $\Omega_2=\Omega_1+\pi$, $\omega_2=\pi-\Omega_1$.}
 we must (i) recognize that S\"oderhjelm's $F$ is minus the Hamiltonian; (ii) multiply $F$ by $n_2$ since $F$ is the Hamiltonian for the fictitious time $f_2$. 

\subsection{\'Cuk \& Burns (2004)}

\nocite{cuk04} \'Cuk \& Burns investigated Brown's Hamiltonian in the context of the motion of the irregular satellites of the giant planets. Here the inner binary consists of the giant planet and the satellite and the outer body is the Sun, so $m_1\ll m_0\ll m_2$.  Since the orbital eccentricities of the outer planets are small, \'Cuk \& Burns restricted themselves to the case of a circular outer binary, $e_2=0$. Then Brown's Hamiltonian is given by equation (\ref{eq:cuk}), and in the notation of \'Cuk \& Burns, this formula should be compared to the Hamiltonian $-(\langle R_E\rangle + \langle R_{E'}\rangle + \langle R_I\rangle)$, where the three terms are defined by their equations (21), (27) and (32). The formulas do not agree. We believe this is because \'Cuk \& Burns neglected the changes in $R_I$ (their eq.\ 13) due to variations in the eccentricity (their eq.\ 19). These variations yield an additional term in the Hamiltonian, which can be written in their notation as
\begin{equation}
\langle R_{I'}\rangle = \tfrac{135}{128}mn_p^2a^2e^2\sqrt{1-e^2}\sin^2i\cos i\cos 2\omega.
\end{equation}
Then we find that $-(\langle R_E\rangle + \langle R_{E'}\rangle + \langle R_I\rangle + \langle R_{I'}\rangle)$ agrees with the second of equations (\ref{eq:cuk}) for Brown's Hamiltonian.

\subsection{Breiter \& Vokrouhlick\'y (2015)}

\cite{bv15} carry out an expansion of the Hamiltonian in which the fictitious time equals the true time or mean anomaly. All of their results are consistent with ours. In particular, the first of our equations (\ref{eq:quadd}) for the double-averaged quadrupole Hamiltonian is equivalent to their equations (49) and (77). The first of our equations (\ref{eq:fordo}) for the double-averaged octopole Hamiltonian is equivalent to their equations (78) and (79). Using computer algebra, it is straightforward to show that their equation (80) for Brown's Hamiltonian is equivalent to our equations (\ref{eq:hbrown}) and (\ref{eq:abmean}) for the Hamiltonian $H_\mathrm{B1}$. 

\subsection{Luo, Katz \& Dong (2016)}

\nocite{luo16} Luo et al.\ evaluate Brown's Hamiltonian using the true anomaly as the fictitious time (they call this analysis ``corrected double averaging'' or CDA). Their equation (39) corresponds to our equations (\ref{eq:hbrown}) and (\ref{eq:abtrue}) for $H_\mathrm{B2}$, after multiplying by the reduced mass of the inner binary $m_0m_1/(m_0+m_1)$ (eq.\ \ref{eq:reduced}). In converting their expression to ours, we eliminate the $y$-components of the dimensionless angular-momentum and eccentricity vectors, using the relations $j_ye_y=-j_xe_x-j_ze_z$ (because $\bj\cdot\be=0$), $j_y^2=1-e^2-j_x^2-j_z^2$ (because $|\bj|^2=1-e^2$) and $e_y^2=e^2-e_x^2-e_z^2$.

Luo et al.'s equation (39) is simplified considerably if it is matched to the Hamiltonian $H_{B3}$ instead of $H_{B1}$. In their notation, (39) becomes 
\begin{equation}
    \Phi=-\epsilon_\mathrm{SA}\frac{27Gm_\mathrm{per}a^2}{64b_\mathrm{per}^3}(1+\tfrac{2}{3}e_\mathrm{per}^2)j_z\big[(1-j_z^2)/3+8e^2-5e_z^2)\big].
\end{equation}

\subsection{Will (2021)}

Will derives and analyzes the equations of motion resulting from Brown's Hamiltonian in terms of orbital elements. Will works directly with the equations of motion, using a two-timescale analysis, and never writes down Brown's Hamiltonian. Nevertheless, using computer algebra it is straightforward to verify that his equations of motion (2.33) are equivalent to those derived from writing Brown's Hamiltonian $H_\mathrm{B1}$ in terms of orbital elements and applying Lagrange's equations for the time evolution of the elements (e.g., \citealt{bc61}, \citealt{md99} or \citealt{tre23}), after (i) rescaling to Will's time coordinate, in which the orbital period of the inner binary is unity; (ii) accounting for his non-standard definition of the longitude of periapsis $\varpi$, which is $\dot\varpi=\dot\omega+\dot\Omega\cos I$ instead of the usual $\dot\varpi=\dot\omega+\dot\Omega,$ and (iii) dividing Brown's Hamiltonian by the reduced mass $m_0m_1/(m_0+m_1)$ to obtain the Hamiltonian per unit mass that is used in Lagrange's equations.

Will's equations of motion correspond to the use of mean anomaly as the fictitious time, so $C(e_2)$ is given by equation (\ref{eq:abmean}). However, he compares his equations of motion to those of \cite{luo16}, who use the true anomaly as the fictitious time (eq.\ \ref{eq:abtrue}). As would be expected from the discussion at the end of \S\ref{sec:q2}, the orbital elements used in these two approaches differ by small oscillatory terms, which Will presents in his Appendix C. With the inclusion of these terms, \cite{luo16} and \cite{will21} are in complete agreement. 

Will's equations of motion (2.33) are simplified considerably if they are derived from the Hamiltonian $H_\mathrm{B3}$ instead of $H_\mathrm{B1}$. The simplified equations are obtained by setting $H(E)=0$ in his equation (2.33). 

\section{Numerical examples}

\label{sec:example}

\begin{figure}
    \includegraphics[width=\columnwidth]{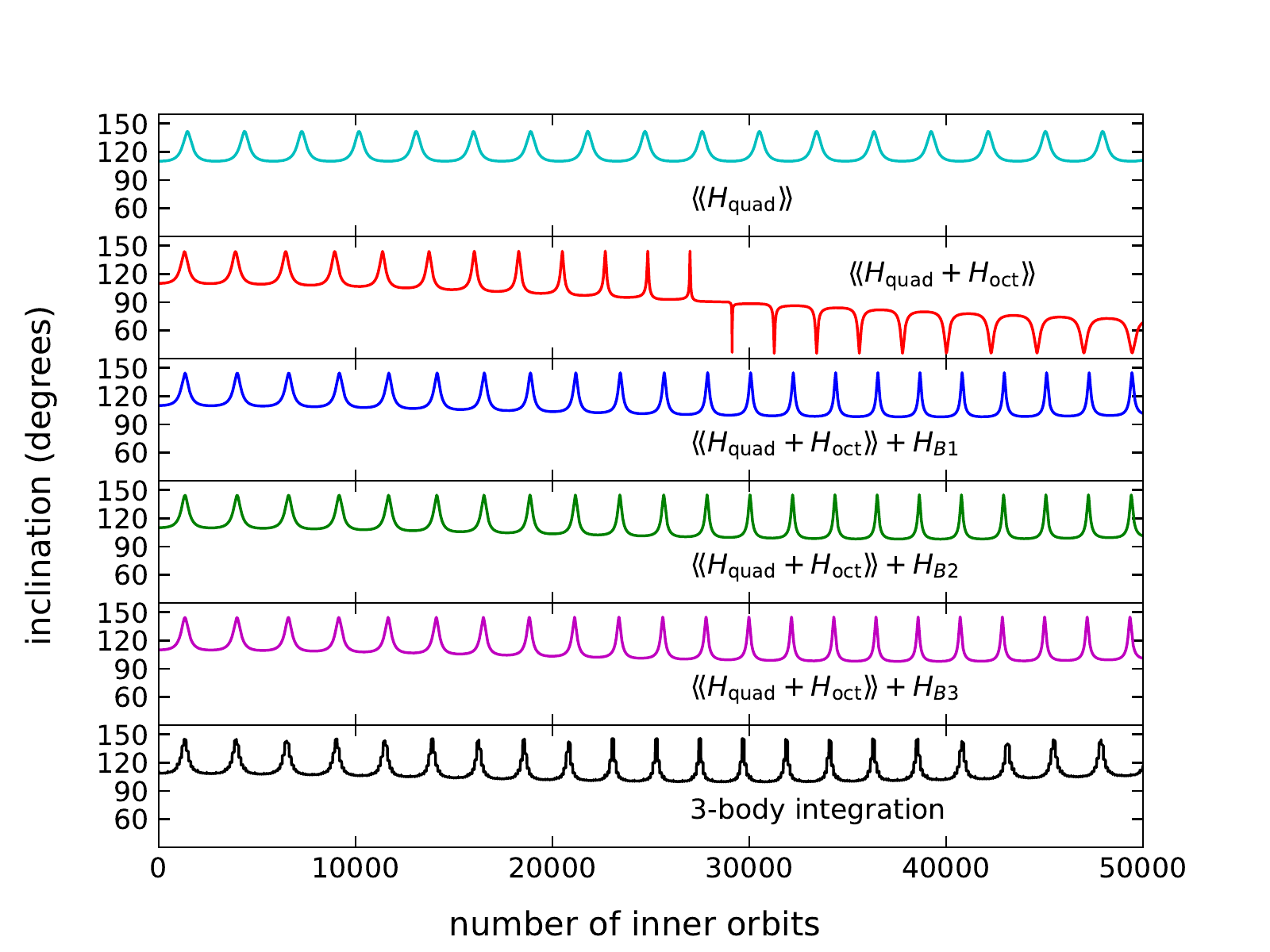}
    \caption{Long-term evolution of a three-body system. The system consists of a test particle orbiting a mass $m_1=1$ at semimajor axis $a_1=1$ with eccentricity $e_1=0.2$. The mass $m_1$ is also orbited by a third body with mass $m_2=1$, semimajor axis $a_2=30$ and eccentricity $e_2=0.8$. Further elements of the orbits are given in the text. The plots show the evolution of the mutual inclination over $50\,000$ initial orbital periods of the inner binary. The bottom panel shows the results of a direct 3-body integration of the equations of motion, while the top four panels show the evolution described by Milankovich's equations (\ref{eq:milank}) using (i) the double-averaged quadrupole Hamiltonian (\ref{eq:quadd}); (ii) the double-averaged quadrupole Hamiltonian plus the double-averaged octopole Hamiltonian (\ref{eq:fordo}); (iii) the double-averaged Hamiltonians plus Brown's Hamiltonian $H_\mathrm{B1}$; (iv) the double-averaged Hamiltonians plus Brown's Hamiltonian $H_\mathrm{B2}$; (v) the double-averaged Hamiltonians plus Brown's Hamiltonian $H_\mathrm{B3}$. All three of Brown's Hamiltonians give solutions that agree well with the 3-body integration.}
\label{fig:one}    
\end{figure}

\begin{figure}
    \includegraphics[width=\columnwidth]{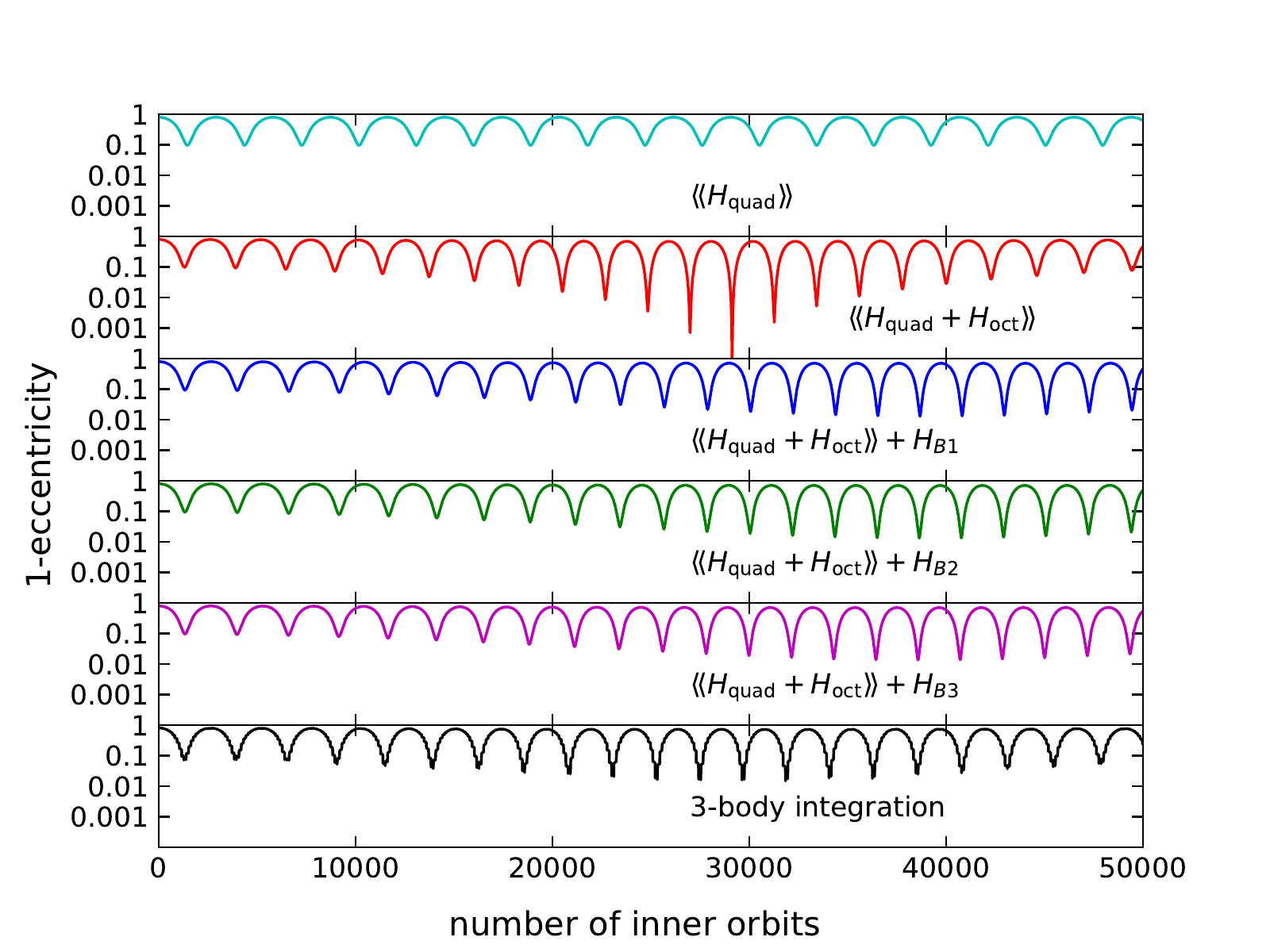}
    \caption{As in Figure \ref{fig:one}, except the eccentricity of the inner orbit is plotted.
    }
\label{fig:two}    
\end{figure}

We have conducted a few numerical experiments to compare the effects of the double-averaged octopole Hamiltonian $\langle H_\mathrm{oct}\rrangle_t$ and the three versions of Brown's Hamiltonian $H_\mathrm{B1}$, $H_\mathrm{B2}$ and $H_\mathrm{B3}$. 

In our first experiment we follow the motion of a test particle orbiting a mass $m_0=1$ that is in turn orbited by a third body, also with mass $m_2=1$. The test particle has initial semimajor axis $a_1=1$ and initial eccentricity $e_1=0.2$. The outer body has semimajor axis $a_2=30$, eccentricity $e_2=0.8$, and orbits in the equatorial plane with its periapsis along the positive $x$-axis. In these coordinates the initial inclination, argument of periapsis, and longitude of the ascending node of the test particle are $I_1=110^\circ$, $\omega_1=0$, and $\Omega_1=\pi$. The test particle is followed for 50\,000 initial orbital periods, both by integrating the Milankovich equations (\ref{eq:milank}) with different Hamiltonians, and by a direct integration of the Newtonian equations of motion using the \textsc{rebound} software package \citep{rein12}. These initial conditions are similar to those in Figure 1 of \cite{luo16}, except that we have chosen a larger eccentricity for the outer body ($e_2=0.8$ instead of 0.2) to emphasize the contribution from the octopole Hamiltonian and the differences between the contributions of the various forms of Brown's Hamiltonian, all of which vanish when $e_2\to 0$. 

The inclinations and eccentricities of the trajectories are shown in Figures \ref{fig:one} and \ref{fig:two}. The bottom panels in each figure show the results of direct integration of the Newtonian 3-body equations of motion. The top panel shows the results of integrating the double-averaged quadrupole Hamiltonian, equation (\ref{eq:quadd}). The inclination and eccentricity undergo regular out-of-phase ZLK oscillations; recall that with this Hamiltonian the $z$-component of angular momentum $j_z=(1-e_1^2)^{1/2}\cos I_1$ is conserved and the motion is integrable. The second panel shows the trajectory when the double-averaged octopole Hamiltonian is added. The motion becomes more complicated and in particular there is an orbital flip (the inclination changes from retrograde to prograde, as the eccentricity nearly reaches unity: $1-e<10^{-4}$) at time $t\simeq 29\,000$.
The following three panels show the effect of adding each of our three expressions for Brown's Hamiltonian, $H_\mathrm{B1}$, $H_\mathrm{B2}$ and $H_\mathrm{B3}$ to the double-averaged quadrupole and octopole potentials. The orbit flip and the large eccentricity oscillation are suppressed; the motion is very similar in all three Hamiltonians and very similar to the direct 3-body integration (apart from a modest cumulative shift in phase in the ZLK oscillations between the 3-body integration and the three integrations using Brown's Hamiltonian).  These findings are consistent with our conclusion that the solutions of the equations of motion for all three forms of Brown's Hamiltonian differ only by small oscillatory terms. 

\begin{figure}
    \includegraphics[trim={0.5cm 0.3cm 1cm 0.5cm},clip,width=\columnwidth]{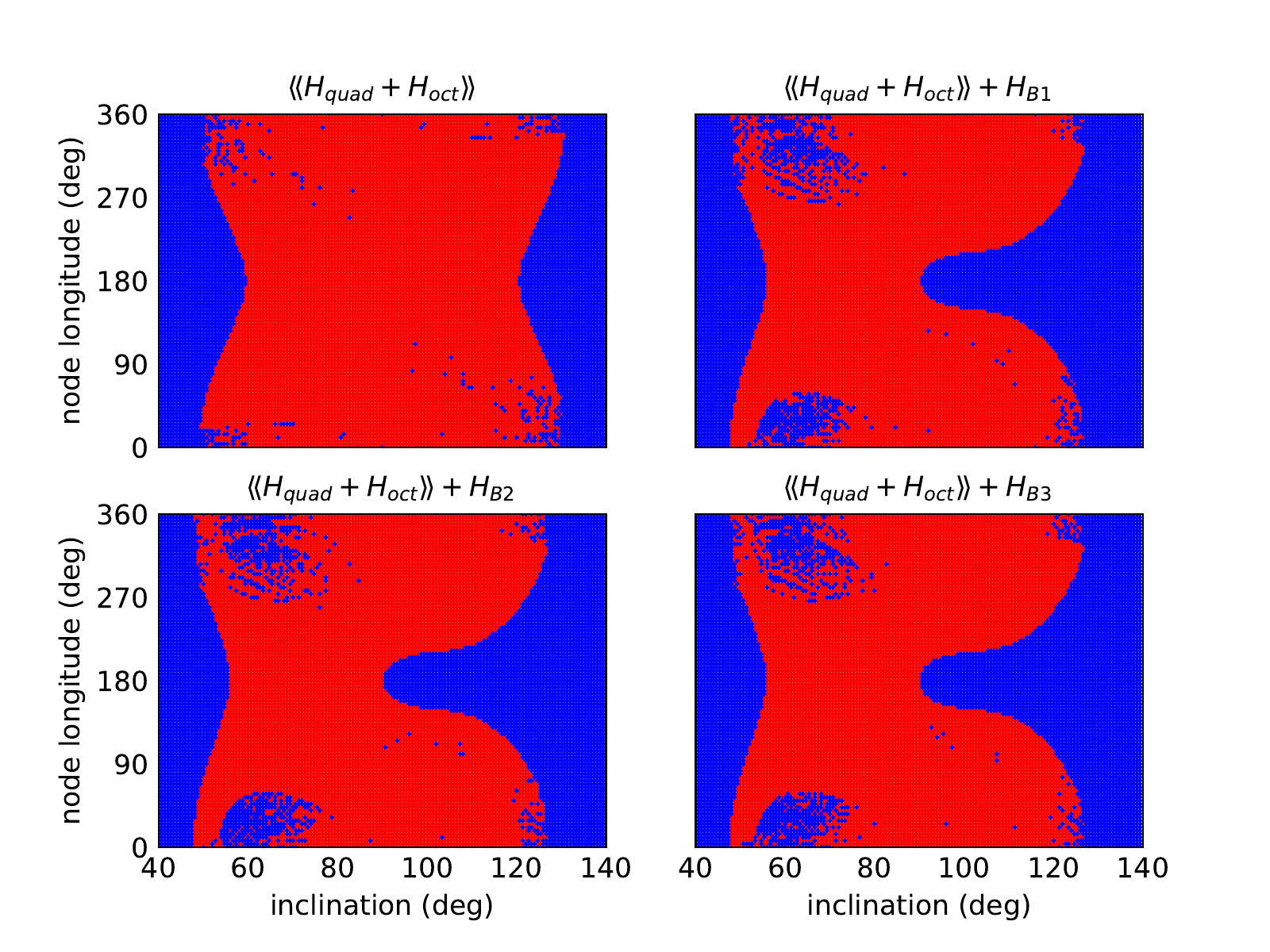}
    \caption{Trajectories with orbital flips are plotted as red points, and those with no flips (constant sign of $\cos I_1$) are blue. The initial orbital elements are the same as in Figures \ref{fig:one} and \ref{fig:two} except that the inclination and longitude of node are scanned over all possible vales. Each orbit is integrated for 125\,000 initial orbital periods. The geography of the orbital flips is essentially the same for all three versions of Brown's Hamiltonian.
    }
\label{fig:three}    
\end{figure}

Figure \ref{fig:three} shows one aspect of the behavior of a family of trajectories. These have the same parameters as in Figures \ref{fig:one} and \ref{fig:two}, except that the inclination varies from $40^\circ$ to $140^\circ$ and the longitude of the ascending node varies from 0 to $360^\circ$. Each trajectory is followed for 125\,000 initial orbital periods, and is plotted as a red point if it flips during the integration ($\cos I_1$ changes sign) and blue if there is no flip. The figure can be directly compared to Figure A6 of \cite{luo16}, which also shows the results for direct 3-body integrations. The main conclusion is that the orbits that exhibit flips are almost the same for all three versions of Brown's Hamiltonian, and for the 3-body integrations. 

\section{Summary}

The classical theory of von Zeipel--Lidov--Kozai or ZLK oscillations in hierarchical triple systems is based on the double-averaged quadrupole potential from a distant body, that is, the potential or Hamiltonian averaged over the orbits of both the inner and the outer bodies. This Hamiltonian is given in equation (\ref{eq:quadd}), both in terms of the dimensionless angular-momentum vector $\bj_1$ and the eccentricity vector $\be_1$ of the inner orbit, and in terms of that orbit's elements $a_1$, $e_1$, $I_1$, $\omega_1$, and $\Omega_1$ (we assume throughout that the angular momentum of the outer orbit is large enough that it remains fixed). The double-averaged quadrupole Hamiltonian conserves the component of the angular momentum of the inner binary normal to the orbit of the outer binary, $j_{1z}=\bj_1\cdot\jhat_2=(1-e_1^2)^{1/2}\cos I_1$, thereby artificially restricting the volume of phase space that the inner binary can access, and in particular prohibiting collisions of the two members of the inner binary unless $j_{1z}\simeq 0$ (if the sum of the radii of the two bodies is much less than their semimajor axis). This shortcoming is usually remedied by including the effects of the double-averaged octopole Hamiltonian (\ref{eq:fordo}) in the equations of motion \citep[e.g.,][]{naoz16}, or by including the variations in the elements of the outer orbit due to perturbations from the inner orbit.  

As described in \S\ref{sec:intro}, when the mass of the outer body is comparable to or larger than the combined mass of the two bodies in the inner binary, $m_2\gtrsim m_0+m_1$, the fractional strength of the double-averaged quadrupole Hamiltonian  relative to the Kepler Hamiltonian for the inner body is $\mbox{O}(\epsilon^2)$ where $\epsilon\sim n_2/n_1$ is the ratio of the mean motions of the outer and inner binary. The fractional strength of the octopole potential, given by equation (\ref{eq:octoom}), can be written as $\mbox{O}(\epsilon^2 a_1/a_2)$, smaller than the quadrupole by $\mbox{O}(a_1/a_2)$.

In comparison, the effects of nonlinear quadrupole perturbations to the orbit of the inner binary, described by Brown's Hamiltonian, are $\mbox{O}(\epsilon^3)$ relative to the Kepler Hamiltonian for the inner body (eq.\ \ref{eq:brownoom}). Therefore these are expected to dominate over the octopole effects whenever 
\begin{equation}
    \frac{m_2}{m_0+m_1}\gtrsim \frac{a_2}{a_1}
\end{equation}
(for example, in the Earth--Moon--Sun system the left side is $3.29\times 10^5$ and the right side is 389). This formula might suggest that in hierarchical systems ($a_2\gg a_1$) with comparable masses ($m_0+m_1\sim m_2$) the effects of Brown's Hamiltonian can be neglected. However, as the examples in \S\ref{sec:example} illustrate, this is not always so.  In practice, we should track the effects of \emph{both} Brown's and the octopole Hamiltonian in almost all astrophysical systems, since (i) the coefficients in the Hamiltonians depending on the orbital elements can vary over a wide range, and typically the coefficients in Brown's Hamiltonian are larger; (ii) the octopole Hamiltonian vanishes in important special cases, such as when the eccentricity of the inner or outer binary is zero or the two masses of the inner binary are the same. 

In some cases one should also include (i) the variations in the outer orbit due to perturbations from the inner binary \citep[e.g.,][]{ford00}; and (ii) the dominant orbit-averaged Hamiltonian arising from general relativistic corrections \citep[e.g.,][]{tre23},
\begin{equation}
    H_\mathrm{gr}=-\frac{3G^4(m_0+m_1)^4}{c^2\Lambda_1^3L_1} + \frac{15G^4(m_0+m_1)^4}{8c^2\Lambda_1^4} + \mbox{O}(c^{-4}),
\end{equation}
in which $c$ is the speed of light and $\Lambda_1$ and $L_1$ are the Delaunay elements defined in equations (\ref{eq:delaunay}).

At least three different forms of Brown's Hamiltonian are found in the literature. We show that these forms arise from different choices of the fictitious time or anomaly that is used for orbit averaging. The presence of these different forms reflects a gauge freedom in the canonical transformations that are used to eliminate secular terms in the equations of motion; the solutions of the equations of motion using the three Hamiltonians differ by small perturbations that oscillate on the timescale of the orbital period of the outer binary. 

In summary, investigations of ZLK oscillations should use the sum of the double-averaged quadrupole and octopole Hamiltonians and Brown's Hamiltonian to characterize the evolution of the triple system. The simplest form of Brown's Hamiltonian is equation (\ref{eq:hb3}) and this is the one that should be used in practice. 

\section*{Acknowledgements}

This work was supported in part by the Natural Sciences and Engineering Research Council of Canada (NSERC), funding reference number RGPIN-2020-03885. I thank the referee, Clifford Will, for discussions that clarified my understanding.

\section*{Data availability}
The codes used in this article will be shared on reasonable request.

\appendix

\section{}

The goal of this Appendix is to show that the $\mbox{O}(\epsilon^3)$ terms in the apsidal and nodal precession rates of the Moon (eq.\ \ref{eq:bc}) can be derived from Brown's Hamiltonian. 

Let the Earth, Moon and Sun be masses $m_0$, $m_1$, $m_2$ respectively. We approximate the solar orbit as circular, $e_2=0$; then since $m_1 \ll m_0 \ll m_2$ Brown's Hamiltonian is given by equation (\ref{eq:cuk}). This can be added to the double-averaged quadrupole Hamiltonian (\ref{eq:quadd}); since the lunar orbit has small eccentricity and inclination, we can expand in powers of $e_1$ and $I_1$, 
\begin{equation}
    \frac{\llangle H_\mathrm{quad}{\rrangle}_t}{m_1}+\frac{H_\mathrm{B}}{m_1}=\frac{3n_2^2a_1^2}{8}(I_1^2-e_1^2) - \frac{9n_2^3a_1^2}{64n_1}(25e_1^2+I_1^2) + \mbox{O}(e_1,I_1)^4, 
\end{equation}
where $n_1$ and $n_2$ are the mean motions of the Moon and Sun. For small eccentricities and inclinations, Lagrange's equations for the rate of precession of the longitudes of periapsis $\varpi_1$ and node $\Omega_1$ read
\begin{align}
\frac{d\varpi_1}{dt}=-\frac{1}{n_1a_1^2e_1}\frac{\p H/m_1}{\p e_1}&= \frac{3n_2^2}{4n_1} + \frac{225n_2^3}{32n_1^2} + \mbox{O}(e_1^2,I_1^2),\nonumber \\
 \frac{d\Omega_1}{dt}=-\frac{1}{n_1a_1^2I_1}\frac{\p H/m_1}{\p I_1}&= -\frac{3n_2^2}{4n_1} +\frac{9n_2^3}{32n_1^2} + \mbox{O}(e_1^2,I_1^2),   
\end{align}
consistent with \nocite{bc61} Brouwer \& Clemence (1961, p.\ 322) and equation (\ref{eq:bc}).

\bsp

\label{lastpage}

\end{document}